\documentclass[aps,twocolumn,amsmath,superscriptaddress]{revtex4}

\usepackage{graphicx}
\usepackage{xspace} 

\usepackage[dvips,bookmarks,a4paper,colorlinks,linkcolor=blue,citecolor=blue,
  menucolor=blue,urlcolor=blue]{hyperref}
\usepackage[all]{hypcap}
\hypersetup{
pdfauthor    = {Rafael B. Dinner},
pdftitle     = {Superconductor-ferromagnet nanocomposites via co-deposition of niobium and dysprosium},
}

\newcommand{\tc}{\ensuremath{T_\mathrm{c}}\xspace}
\newcommand{\jc}{\ensuremath{J_\mathrm{c}}\xspace}

\newcommand{\hctwo}{\ensuremath{H_\mathrm{c2}}\xspace}

\newcommand{\Tc}{\tc{}}

\newcommand{\dg}{\ensuremath{^\circ}\xspace}

\begin{document}

\title{Superconductor-ferromagnet nanocomposites via co-deposition of niobium and dysprosium}
\author{Rafael B. Dinner}
\email[Electronic address: ]{rd356@cam.ac.uk}
\author{Suman-Lata Sahonta}
\affiliation{Department of Materials Science, Cambridge University, Pembroke Street, Cambridge CB2 3QZ, United Kingdom}
\author{Rong Yu}
\affiliation{Department of Materials Science and Engineering, Tsinghua University, 100084 Beijing, People's Republic of China}
\author{Nadia A. Stelmashenko}
\author{Judith L. MacManus-Driscoll}
\author{Mark G. Blamire}
\affiliation{Department of Materials Science, Cambridge University, Pembroke Street, Cambridge CB2 3QZ, United Kingdom}

\date{\today}

\begin{abstract}
We have created superconductor-ferromagnet composite films in order to test the enhancement of critical current density, \jc{}, due to magnetic pinning. We co-sputter the type-II superconductor niobium (Nb) and the low-temperature ferromagnet dysprosium (Dy) onto a heated substrate; the immiscibility of the two materials leads to a phase-separated composite of magnetic regions within a superconducting matrix. Over a range of compositions and substrate temperatures, we achieve phase separation on scales from 5~nm to 1~$\mu$m. The composite films exhibit simultaneous superconductivity and ferromagnetism. Transport measurements show that while the self-field \jc{} is reduced in the composites, the in-field \jc{} is greatly enhanced up to the 3~T saturation field of Dy. In one instance, the phase separation orders into stripes, leading to in-plane anisotropy in \jc{}.
\end{abstract}


\maketitle

\begin{flushleft}
Note: The latest version of this paper with higher quality graphics is available from \texttt{\href{http://alum.mit.edu/www/rdinner/research}{http://alum.mit.edu/www/rdinner/research}}.

It should be referenced as R. B. Dinner \textit{et al.}, \href{http://dx.doi.org/10.1088/0953-2048/22/7/075001}{\textit{Supercond. Sci. Technol.} \textbf{22} (2009) 075001}.
\end{flushleft}

\section{\label{intro}Introduction}

Achieving high critical current densities, \jc{}, in superconductors requires effective vortex pinning. Pinning occurs naturally via material inhomogeneity and microstructural defects, but can be enhanced through the addition of artificial pinning centres, such as insulating particles within a YBCO matrix \cite{HTS_review,bzo,tantalates}. Here we study whether \textit{magnetic} artificial pinning centres can further improve \jc{}. A non-magnetic, insulating cylinder of radius $\xi$ (the coherence length of the superconductor) aligned with a vortex already appears optimal in that the full condensation energy of the vortex core is gained---leading to the strongest core pinning force---with a minimum loss of superconducting material. Yet many have suggested that the magnetic portion of the free energy of a vortex offers another route to strengthen pinning \cite{SF_hybrids,cardoso,milosevic}. This premise was the basis for several experiments in which arrays of magnetic dots were created on top of a superconducting film, or a film was deposited over such an array \cite{velez_review,snezhko,lange,morgan}.

In these previous experiments, the interaction between vortex and magnet falls into two categories, neither of which offers superior pinning to an ideal core pin. In the first, a vortex interacts with the stray field of a magnetic dot. The resulting potential well for the vortex may be deep, but its width is governed by the penetration depth $\lambda$, which in practical superconductors is relatively large compared to $\xi$, leading to a low pinning \textit{force}. The other interaction is the proximity effect, in which the magnet suppresses superconductivity, which also cannot improve upon an ideal core pin, where superconductivity is suppressed completely. Blamire \textit{et al.} \cite{Blamire_magnetic_pinning_theory} therefore proposed that the greatest \jc{} enhancement will arise not from altering the potential landscape of vortices, but from reducing their magnetic flux and thereby reducing the Lorentz force acting on those vortices in the presence of a transport current. However, magnetic dots above or below a superconducting film, as in the experimental work referenced above \cite{velez_review,snezhko,lange,morgan}, do not achieve this Lorentz force reduction; within the superconductor the vortices retain their full flux. Instead, the magnetic material must be embedded in the film, ideally passing all the way through it with vortices lying entirely along the inclusions. Furthermore, the inclusions must have sufficient magnetisation to divert a significant portion of a flux quantum, which was not the case for several experiments with small ($\sim5$~nm) inclusions \cite{NbGd, Rizzo, Co_clusters_in_Pb}. Ref.~\cite{Rizzo}, in particular, shows that magnetic inclusions remain effective at smaller sizes than non-magnetic inclusions, but because the mechanism is still core pinning rather than flux reduction, the maximum pinning force does not improve. In this work, we instead seek to optimise flux diversion by creating larger, strongly magnetic pinning centres within a niobium film.

We take a simple approach to creating a superconductor with embedded magnetic material: we co-deposit its constituent elements. For most choices of materials, however, this would lead to atomic-scale mixing (alloying), which would degrade both the superconductivity and the magnetism. In particular, $s$-wave Cooper pairs are highly sensitive to magnetic scattering, requiring regions of magnet-free superconductor larger than the superconducting coherence length \cite{magnetism_suppresses_tc}. So we have chosen the superconductor niobium (Nb) and the rare earth low-temperature ferromagnet dysprosium (Dy) in the hope that they do not alloy. Although, to our knowledge, this system has not been investigated previously, a chemically similar system of tantalum (Ta) and Dy was found to be immiscible \cite{TaDy}. During growth, if the atoms have sufficient mobility, the film should phase-separate into the desired regions of pure superconductor and pure ferromagnet. Dy exhibits a very high saturation magnetisation of 3.7~T below its Curie temperature of 88~K, thus a 30~nm diameter inclusion could carry an entire flux quantum, $2\times10^{-15}$~Wb. By growing such inclusions, ideally in a columnar structure, we aim to strongly divert magnetic flux from vortices.

In the following sections, we show that niobium and dysprosium do phase-separate, exhibiting a variety of microstructures as a function of the film composition and growth temperature. We characterise the microstructures with X-ray diffraction, atomic force microscopy (AFM), and transmission electron microscopy (TEM) coupled with element-sensitive detection techniques. We then examine the superconducting properties with both magnetisation and transport measurements. We find coexistence of superconductivity and magnetism over most of the growth parameter space, and enhanced critical current in magnetic fields up to the saturation field of the dysprosium.

\section{Film growth}

The films are approximately 200~nm thick, grown by dc magnetron sputtering. Niobium and dysprosium targets are co-sputtered to deposit a mixture onto heated R-plane $[1\bar{1}02]$ sapphire (Al$_2$O$_3$) substrates. The composition is controlled via the relative magnetron powers to have atomic ratios of Nb:Dy 1:1, 2:1, and 4:1, with a total deposition rate of approximately 5~nm/min. At a fixed film composition, films with a range of substrate temperatures are grown in each sputtering run. The temperatures reported are those of the heater segment directly under each substrate; the samples are $\sim150$~K cooler, based on a prior calibration. Liquid nitrogen is applied to the walls of the deposition chamber to achieve a base pressure of $4\times10^{-8}$~Pa. The substrate heater raises this pressure, but baking out the heater keeps the deposition background below $1\times10^{-7}$~Pa. For sputtering, the chamber is then filled to 0.9 Pa with high-purity argon.

Before depositing the Nb:Dy mixture, a third magnetron is used to deposit a 50~nm thick buffer layer of Ta in order to preclude any possibility of a reaction between the film and the substrate. In particular, a rare earth such as Dy may react with the oxygen in Al$_2$O$_3$, freeing Al to combine with Nb \cite{Nb_buffer_layer}. Nb$_3$Al is a superconductor with a large upper critical field (\hctwo{}) that could contribute to the transport properties of the film and distract from our analysis of the composite. The Ta buffer layer does not significantly affect the transport or magnetic behaviour at the measurement temperature of 4~K, as confirmed by a control sample consisting of the tantalum layer only. A set of Nb:Dy films with a 100~nm Ta layer was also grown to check that the behaviour was unchanged from those with a 50~nm Ta layer, indicating that there is no reaction between the film and the substrate mediated by the buffer thickness.

\section{Microstructural characterisation}
\label{microstructure}

\begin{figure}
  \includegraphics[width=86mm]{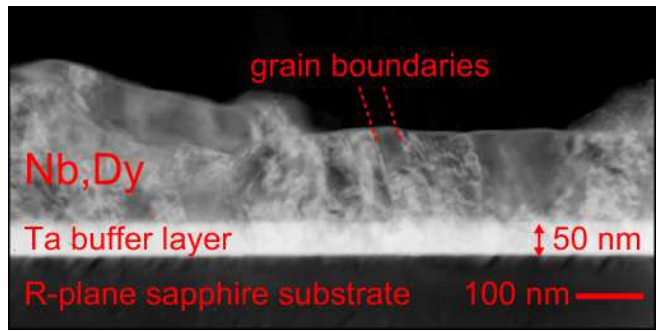}
  \caption{\label{fig_xsec_tem}Cross sectional TEM image of a Nb:Dy 2:1 film grown at 760\dg{}C shows some nearly vertical grain boundaries, but a predominance of disordered growth.}
\end{figure}

X-ray diffraction reveals that both niobium and dysprosium exhibit crystalline phases within the composite, and some degree of orientation. The width of the peaks, however, indicates that the grains are smaller than the 200~nm thickness of the film. The cubic niobium grows [001] oriented; the rocking curve of the [002] peak has a full width half maximum (FWHM) of 1.0 to 1.3\dg{}, while the hexagonal dysprosium shows a [101] peak with a rocking curve FWHM of 3\dg{}.

Cross-sectional TEM indicates some areas of columnar growth, in the form of tilted grain boundaries, seen in the bright-field image, Fig.~\ref{fig_xsec_tem}. Overall, however, the growth is not highly ordered, with the TEM confirming the x-ray result that the grain size is much smaller than the 200~nm thickness of the film in many places. Energy filtering did not yield sharp contrast between regions of Nb and Dy, suggesting that the two overlap at a length scale below the $\sim50$~nm thickness of the cross-sectional specimen. Thus we expect a vortex in the material to pass through regions of Dy and Nb, rather than being optimally pinned by a through-thickness column of Dy.

\begin{figure*}
  \includegraphics[width=177mm]{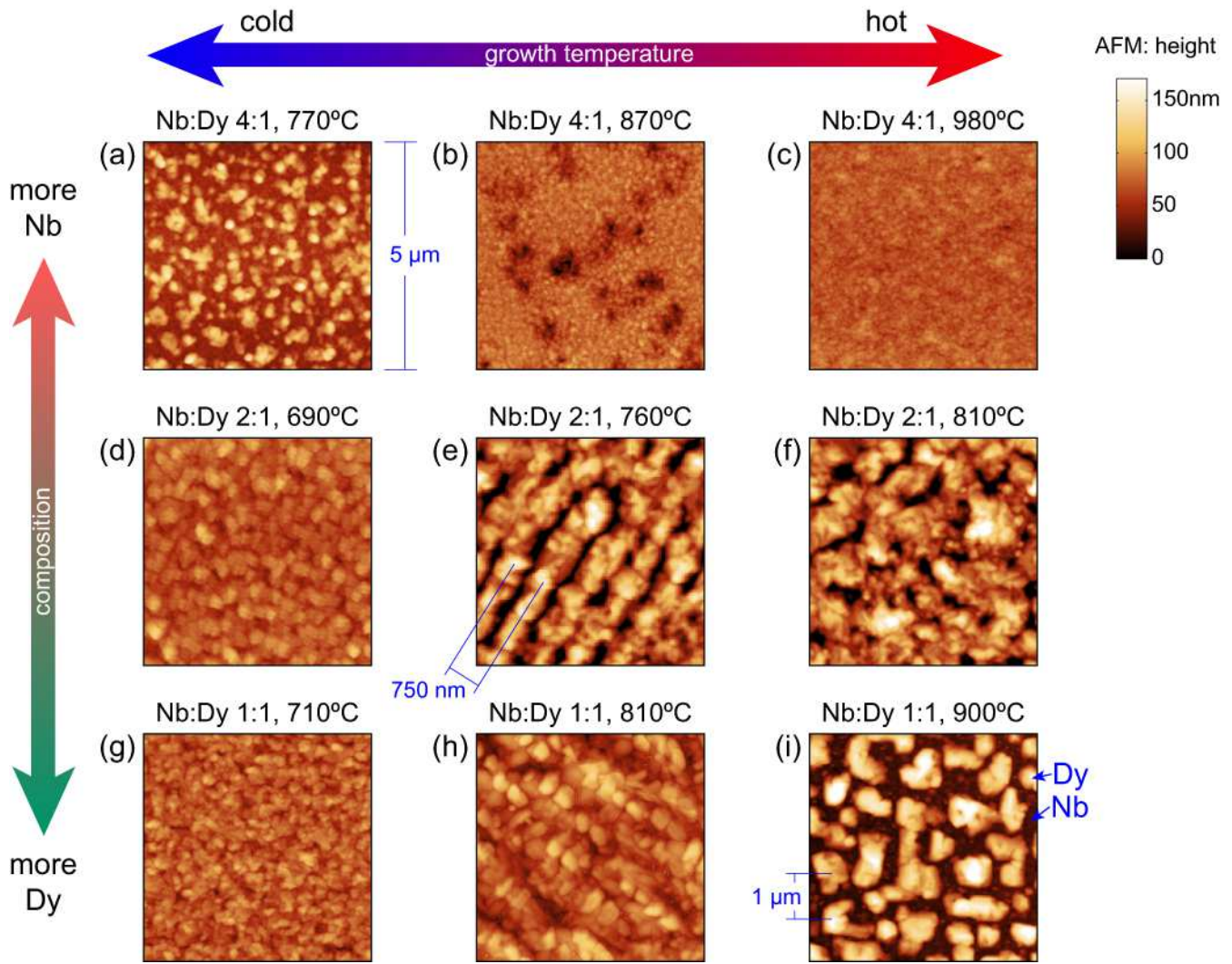}
  \caption{\label{fig_afm}Atomic force microscopy (AFM) images of composite films, illustrating the variety of microstructures accessible via various growth conditions. All images are on the same height and length scales.}
\end{figure*}

The thickness variation seen in the cross-sectional TEM is not an artefact; the AFM images in Fig.~\ref{fig_afm} show that some films have thickness corrugations of more than 100~nm. This corrugation is due to phase separation, as explained below, and we expect higher growth temperatures, corresponding to higher mobility for atoms on the deposition surface, to yield a larger length scale for this separation. This holds true for the samples with the highest concentration of Dy, Fig.~\ref{fig_afm}(g)--(i), where the length scale of the corrugation increases with increasing growth temperature. By examining these three samples in a scanning electron microscope (SEM), we find that the topography (from secondary electrons) correlates with composition, measured with energy dispersive spectrometry (EDS) maps of Nb and Dy concentration. For the film in Fig.~\ref{fig_afm}(i), EDS indicates that the high islands are at least 90 atomic \% dysprosium and the surrounding low matrix is at least 90\% niobium. The films in (h) and (g) show successively less composition contrast, probably because the length scale of separation is becoming smaller than the resolution of the EDS (1~$\mu$m).

In the Nb:Dy 4:1 samples, in contrast, the length scale of corrugation decreases with increasing growth temperature (Fig.~\ref{fig_afm}(a)--(c)). The trend is borne out by the six samples grown, of which three are shown. So while the growth temperature has different effects for different compositions, it does offer a degree of control over the microstructure. In Section~\ref{section_jc}, we show that this in turn has a dramatic effect on the superconducting performance, with the high temperature, fine-grained films yielding the best in-field \jc{}.

An interesting stripe texture is visible in the topography of two of the films, Fig.~\ref{fig_afm}(e) and (h). The stripes also show composition contrast in SEM EDS maps. They are are aligned to the [100] axis of Nb (so within the $[1\bar{1}02]$ R-plane of the sapphire, they are 45\dg{} from the [0010] axis), suggesting an easy diffusion direction during growth. This is plausible, since the [001] Nb is known to grow $\sim3$\dg{} misoriented relative to the R-plane, potentially leading to the stepped surface typical of a vicinal film \cite{Nb_on_sapphire2}. A stepped surface can also arise from miscut R-plane sapphire \cite{Nb_on_sapphire2}, though we have not applied the temperatures typically necessary to develop such steps, and do not observe them on our substrates. Regardless of their origin, we expect the stripes' compositional and topographic variation to produce in-plane anisotropy in the transport properties of the film, which is indeed observed, as described in Section~\ref{stripe_jc}.

\begin{figure*}
  \includegraphics[width=177mm]{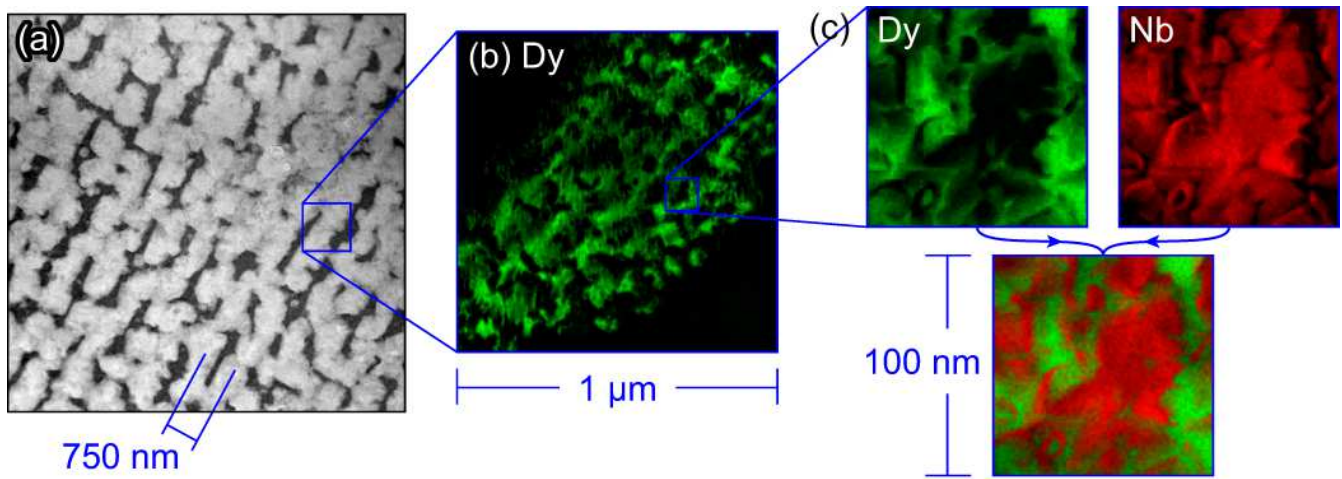}
  \caption{\label{fig_tem_zoom}Plan-view transmission electron microscopy (TEM) images of the Nb:Dy 2:1 film grown at 760\dg{}C. (a) High-angle annular dark field Z-contrast image. The dark regions are pure Nb and the light regions are a mixture of Nb and Dy. (b) Energy-filtered TEM image in which Dy appears bright. (c) Energy dispersive spectrometry (EDS) mapping shows that Nb and Dy fully phase-separate, but are interwoven at length scales down to 5~nm.}
\end{figure*}

To examine the phase separation with higher resolution, we turn to TEM, looking in plan-view at this same striped film. The stripes are again visible in a high-angle annular dark field (HAADF) image (Fig.~\ref{fig_tem_zoom}(a)), where the higher atomic number of Dy causes more scattering, appearing bright compared to Nb. High magnification TEM reveals that the phase separation has additional structure on a much smaller scale. The energy-filtered image of Dy regions in Fig.~\ref{fig_tem_zoom}(b) confirms that the dark regions in (a) are pure Nb, while the light stripes are a mixture of Nb and Dy. Zooming in further on a light stripe in (c) reveals complete phase separation only at a 5~nm length scale. Here, we use EDS to quantitatively map the concentration of each element, displayed as the brightness of two colours, green for Dy and red for Nb. Overlaying the images, we can see red and green regions, but no yellow regions in which both are present, so Nb and Dy are fully phase-separated.

\section{Coexistence of magnetism and superconductivity}

\begin{figure}
  \includegraphics[width=86mm]{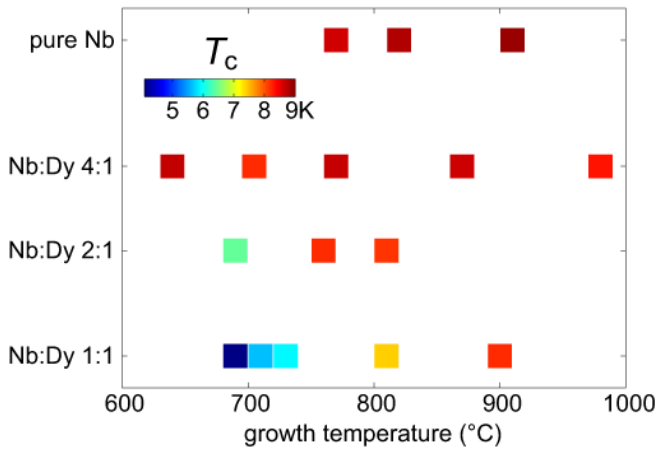}
  \caption{\label{fig_tc}The transition temperature of the composite films, measured by dc transport, does not degrade substantially until we lower the niobium content to 50~atomic percent and decrease the growth temperature to 700\dg{}C.}
\end{figure}

A typical problem with incorporating magnetic materials into superconductors is poisoning of the superconductivity, manifested by a drop in \tc{} and leading to suppression of \jc{} \cite{magnetism_suppresses_tc}. We do see a suppression of \tc{} at high concentrations of Dy and low growth temperatures, as seen in Fig.~\ref{fig_tc}, where we plot transition temperatures obtained by dc transport measurements, with \tc{} identified as the point of maximum slope in resistance versus temperature. The small-scale phase separation observed in TEM (described above) can explain the reduced \tc{}. We expect that mixing at a length scale below the 20~nm coherence length of pure Nb would suppress the order parameter in the Nb regions due to the proximity effect \cite{proximity_suppressed_tc}. Fortunately, however, \tc{} is preserved in our films over most of the composition-growth temperature phase diagram, suggesting that the Nb and Dy do fully phase-separate at larger length scales.

\begin{figure}
  \includegraphics[width=86mm]{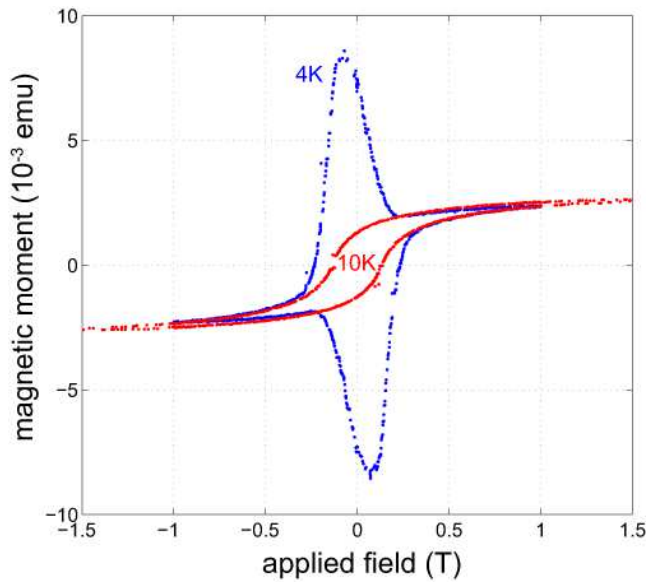}
  \caption{\label{fig_vsm}Coexistence of superconductivity and ferromagnetism in the Nb:Dy 2:1 film grown at 760\dg{}C. In this vibrating sample magnetometer (VSM) measurement, the sample is oriented nearly parallel to the field in order to reduce the magnitude of the diamagnetic superconducting response to make it comparable to the ferromagnetic response.}
\end{figure}

It is perhaps not surprising, then, that magnetisation measurements clearly show the coexistence of superconductivity and magnetism, seen in Fig.~\ref{fig_vsm}. These measurements were performed with a cryogenic vibrating sample magnetometer (VSM). At 10~K, above \tc{}, the ferromagnetic hysteresis loop of the Dy is the only response visible. Below \tc{} at 4~K, superconducting diamagnetism simply superposes on the ferromagnetism. The film is oriented approximately parallel to the field to make its superconducting response, which would otherwise dominate, comparable in magnitude to that of the ferromagnet.

\section{Enhanced critical current in field}
\label{section_jc}

\begin{figure*}
  \includegraphics[width=177mm]{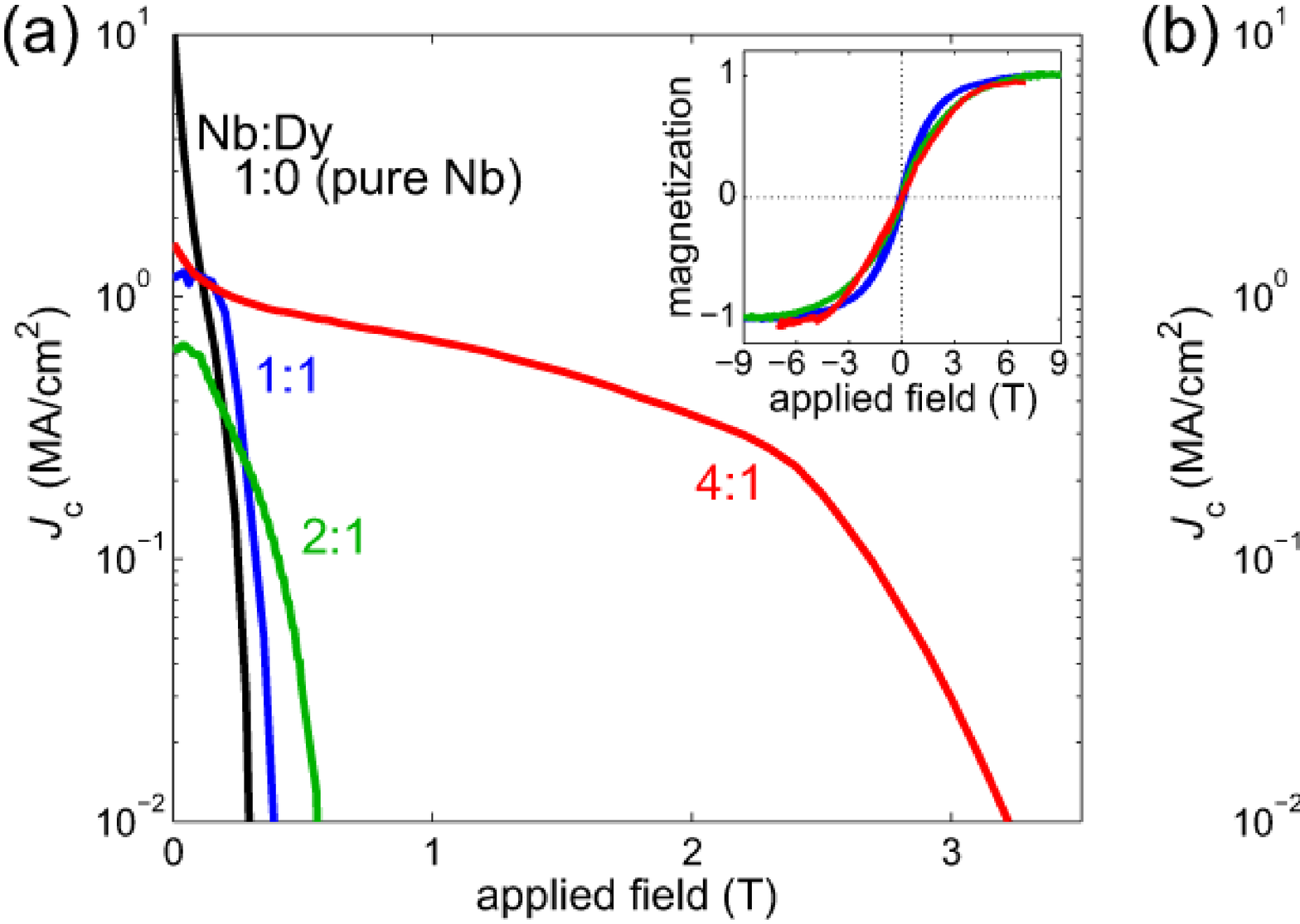}
  \caption{\label{fig_jc}(a) Critical current versus applied magnetic field at 4~K of composite films with compositions as labeled. The optimum growth temperature is shown at each composition: Nb:Dy 1:1 900\dg{}C, Nb:Dy 2:1 810\dg{}C, and Nb:Dy 4:1 925\dg{}C. A pure Nb film grown at 910\dg{}C is measured for comparison. The inset shows, in corresponding colors, the magnetisation of the same films above \tc{} at 10~K, normalised by their saturation magnetisations. (b) The evolution of \jc{} with growth temperature in Nb:Dy 4:1 films. The in-field performance improves up to 920\dg{}C.}
\end{figure*}

We gauge the performance of the composites using transport measurements of \jc{}. The films are patterned into 50~$\mu$m wide (10~$\mu$m for the stripe sample) by 0.5~mm long bridges with a standard 4-point geometry, using photolithography and argon ion milling. The magnetic field is applied perpendicular to the film surface, and \jc{} is determined with a voltage criterion of 0.1~$\mu$V. For the data shown, the samples are cooled to 4~K by static helium exchange gas. Data was also taken at temperatures from 3 to 8~K, but yielded similar results. After accounting for the residual field applied by our magnet, \jc{} does not display an offset or hysteresis in the zero-field peak greater than 10~mT when the field was reversed, as might be expected from reverse domain superconductivity \cite{rds}, so we show only one sign of applied field.

The composite films exhibit remarkably high critical currents at high fields. Fig.~\ref{fig_jc}(a) shows that the pure Nb displays the highest self-field \jc{}, but the composites outperform it for fields above 0.2~T. While the self-field \jc{} among the composites does not show a clear trend with composition, the in-field performance improves dramatically with decreasing Dy content over the composition range that we have explored (Fig.~\ref{fig_jc}(a)). It is notable that the optimum field at which high \jc{} persists in the composites is the field at which the magnetisation of dysprosium begins to saturate (see the inset of Fig.~\ref{fig_jc}(a)), consistent with the prediction of Blamire \textit{et al} \cite{Blamire_magnetic_pinning_theory} that above their saturation field, the inclusions cannot shield the superconductor from flux as effectively. 

The low-field analysis in Ref.~\cite{Blamire_magnetic_pinning_theory} suggests, though, that the greatest \jc{} improvement should occur at zero field, which is not supported by our data. We can instead suppose that the inclusions, by diverting magnetic flux, cause the superconducting matrix to see a decreased field, rather than a decreased flux of individual vortices. The average field $\bar{B}$ is then the sum of this decreased field, $B'$, and the magnetisation of the inclusions, $M$:
\begin{equation}
  \label{B_transform}
	\bar{B} = B' + \mu_0 M(B') \cdot f
\end{equation}
where $f$ is the volume fraction of magnetic material; this notation is consistent with Ref.~\cite{Blamire_magnetic_pinning_theory}. To compare \jc{} of the plain Nb and Nb:Dy samples, we use Eq.~\ref{B_transform} to scale the field values in the plain Nb data from $B'$ to $\bar{B}$, having measured the normal state magnetisation response $M(B')$ for each composite (seen in the inset of Fig.~\ref{fig_jc}(a)) and $f$, the volume fraction of Dy. This shifts \jc{} values to higher applied fields with the addition of magnetic material, as observed for our composites. Comparing the scaled Nb data with the composites' data, we find that while the scaling can explain the performance of the 2:1 Nb:Dy film, the scaling decreases with decreasing Dy content, and therefore overpredicts the in-field performance of the 1:1 composite, and underpredicts the 2--3~T field at which the 4:1 film maintains a substantial \jc{}.

We must therefore conclude that \jc{} is affected by the Dy in other ways, not just field screening. For example, it is plausible that the addition of Dy also causes microstructural defects that decrease the mean free path, decreasing the coherence length, and increasing the upper critical field \hctwo{}, thereby allowing \jc{} to persist up to higher fields than in pure Nb \cite{Hc2_enhancement}. The temperature dependence of \jc{} in the best films (Nb:Dy 4:1) supports the notion that \hctwo{} limits \jc{}: The field $H(T)$ at which \jc{} drops off can be fit to $H(T)=H(0)(1-(T/T_c)^2)$ where $T$ is the temperature. This is the functional form for the phenomenological temperature dependence of \hctwo{} \cite{Poole}.

While the temperature dependence for a given film follows \hctwo{}, the differences between the composites cannot be entirely explained by this parameter. For a system in the dirty limit, \hctwo{} is roughly inversely proportional to the residual resistivity ratio (RRR), resistivity at room temperature divided by resistivity just above \tc{}. The 2:1 and 4:1 composites both have RRRs of 5 compared to 14 for the pure Nb sample, thus we expect \hctwo{} to be higher in these films, though this effect, like the flux shielding, underpredicts the performance of the 4:1 film, and cannot explain why 4:1 is so much better than 2:1. \hctwo{} enhancement, then, is likely an important factor for the in-field \jc{} of the composites, but must be coupled with other pinning enhancements to produce the high \jc{} of the 4:1 material.

As further evidence for other factors at work, the optimised \jc{} exhibits a strong dependence on growth temperature. Results are shown for the best-performing composition of Nb:Dy 4:1 in Fig.~\ref{fig_jc}(b). The trend with growth temperature is clear: as the temperature increases from 700 to 920\dg{}C, the critical field increases at the expense of a decreasing self-field \jc{}. Above 920\dg{}C, \jc{} drops throughout the field range. At high field, then, there is an optimum growth temperature around 920\dg{}C. The RRRs for these films (ranging from 4--6) cannot explain their varying performance, thus it cannot reflect a simple \hctwo{} increase. It is instead likely that the smaller grain size of the 4:1 films grown at high temperature (see Section~\ref{microstructure}) is increasing \jc{}, as has been observed for pure Nb~\cite{Nb_GB_pinning}.

\section{Anisotropic transport from stripes}
\label{stripe_jc}

\begin{figure}
  \includegraphics[width=86mm]{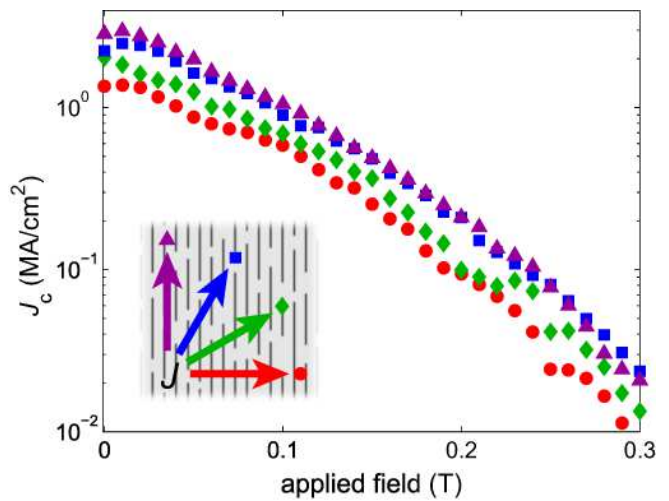}
  \caption{\label{fig_jc_vs_phi}Critical current, \jc{}, versus applied magnetic field at 4~K for the Nb:Dy 2:1 film grown at 760\dg{}C. The colours and symbols indicate the in-plane angles of the four bridges measured, as shown in the inset. \jc{} is highest parallel to the stripes present in the topography and composition of the film (see Fig.~\ref{fig_tem_zoom}).}
\end{figure}

The film that exhibits the stripes in topography and composition shown in Fig.~\ref{fig_tem_zoom} (Nb:Dy 2:1 grown at 760\dg{}) also exhibits in-plane anisotropy in \jc{}. By patterning bridges at a series of in-plane angles (see Fig.~\ref{fig_jc_vs_phi}), we find that \jc{} is approximately twice as high when the current is applied parallel to the stripes as perpendicular, with the intermediate angles following the trend. This can be explained simply by the topographic variation visible in Fig.~\ref{fig_afm}(e): current traversing the stripes must cross regions half as thick as the rest of the film, whereas current running along the stripes can, to a large extent, avoid these regions. Such anisotropic transport structures of interconnected filaments are desirable in superconducting power applications to limit ac losses \cite{ac_losses,ac_losses2}. Not only does this structure emerge naturally in our system, but it does so with very small filament pitch (750~nm), which should yield proportionally smaller ac losses compared to artificially patterned films \cite{ac_losses2}.

\section{Conclusions}

Co-deposition of immiscible metals has proven to be a simple but successful method of producing a superconductor-ferromagnet nanocomposite that vastly outperforms the pure superconducting material. We find that for less than 50\% Dy or growth temperatures above 750\dg{}C, \Tc{} is not suppressed, and superconductivity coexists with magnetism, indicating full phase separation. The magnetic inclusions lead to enhanced \jc{} above 0.2~T, up to their saturation field of 3~T. This performance is sensitive to both composition and growth temperature.

The high-field performance of the optimum composite material cannot be fully accounted for with calculated improvements due to the magnetism of the Dy and the enhanced \hctwo{} of the material. So it remains likely that microstructural changes in the composites are also essential to the observed pinning enhancement. Because the microstructure is not well controlled, though, we cannot construct comparable non-magnetic composites to isolate the contribution of magnetism to \jc{}. Nonetheless, the coincidence of the saturation field of Dy and the field range of enhanced \jc{} is evidence for the role of high magnetic susceptibility in the pinning. Better control of the nanoscale structure of the pinning centres in future work may reveal the full potential of this magnetic pinning.

\begin{acknowledgments}
We thank Drs~Yury Bugoslavsky and Mannan Ali for preliminary magnetometry, and Drs~Stuart Wimbush and John Durrell for measurement assistance. R.D. thanks Hughes Hall, Cambridge for a research fellowship. This work was supported by the UK Engineering and Physical Sciences Research Council [grant number E011020].
\end{acknowledgments}

\providecommand{\newblock}{}

\end{document}